\documentclass{aa}  

\usepackage{graphicx}
\usepackage{txfonts}
\usepackage{natbib}

\begin{document}

   \title{Comparing the birth rate of stellar black holes \\ 
   in binary black hole mergers and long GRBs
   }
   \subtitle{}
\titlerunning{The birth rate of stellar black holes}

   \author{
      J.-L. Atteia\inst{1}
         \and
      J.-P. Dezalay\inst{1}
         \and
      O. Godet\inst{1}
         \and
      A. Klotz\inst{1}
         \and
      D. Turpin\inst{1}
         \and
      M. G. Bernardini\inst{2,3}
%         \and
%      F. Piron\inst{2}
     }

   \institute{
      IRAP, Universit\'e de Toulouse, CNRS, UPS, CNES, Toulouse, France\\ 
      \email{jean-luc.atteia@irap.omp.eu}
         \and
      Laboratoire Univers et Particules de Montpellier, Universit\'e de Montpellier, CNRS/IN2P3, Montpellier, France
         \and
      INAF-Osservatorio Astronomico di Brera, via E. Bianchi 46, I-23807 Merate, Italy
   }

   \date{Received ...; accepted ...}

% \abstract{}{}{}{}{} 
% 5 {} token are mandatory
 
  \abstract
   {Gravitational wave interferometers have proved the existence of a new class of binary black holes (BBHs) weighting tens of solar masses, and they have provided the first reliable measurement of the rate of coalescing black holes (BHs) in the local universe.
   On another side, long gamma-ray bursts (GRBs) detected with gamma-ray satellites are believed to be associated with the birth of stellar mass black holes, providing a measure of the rate of these events across the history of the universe, thanks to the measure of their cosmological redshift.
   These two types of sources, which are subject to different detection biases and involve BHs born in different environments with potentially different characteristics, provide complementary information on the birth rate of stellar black holes.}
%   Under reasonable assumptions concerning the delay before the coalescence of the system, this rate can be used to calculate the birth rate of massive stellar black holes in binary systems.}
   {We compare the birth rates of black holes found in binary black hole mergers and in long gamma-ray bursts.}
   {We construct a simple model which makes reasonable assumptions on the history of GRB formation, and which takes into account some major uncertainties, like the beaming angle of GRBs or the delay between the formation of BBHs and their coalescence.
   We use this model to evaluate the ratio of the number of stellar mass BHs formed in BBH mergers to those formed in GRBs.}
   {We find that in our reference model the birth rate of stellar black holes in BBH mergers represents from few percent to 100\% of the rate of long GRBs and that comparable birth rates are favored by models with moderate beaming angles.
   These numbers, however, do not consider sub-luminous GRBs, which may represent another population of sources associated with the birth of stellar mass black holes.
   We briefly discuss this result in view of our understanding of the progenitors of GRBs and BBH mergers, and we emphasize that this ratio, which will be better constrained in the coming years, can be directly compared with the prediction of stellar evolution models if a single model is used to produce GRBs and of BBH mergers with the same assumptions.}
   %a ratio that is compatible with the hypothesis that every BH in a BBH merger is born with a GRB.
   %We discuss some consequences of this finding in the context of stellar BH formation.}
   %under the two opposite hypotheses that all BHs formed in binary systems produce a GRB or that they produce no GRB.}
   {}


   \maketitle
%
%-------------------------------------------------------------------

\section{Introduction}
\label{sec_intro}
The LIGO and VIRGO interferometers have opened a new window on the universe with the detection of gravitational waves (GWs) from the coalescence of massive stellar black holes in binary systems \citep{Abbott2016c,Abbott2016a,Abbott2017a}. 
Beyond the performance of detecting gravitational waves on the Earth, these observations had a strong astronomical impact with the detection of previously unknown binary systems of two massive stellar black holes.
Additionally, the gravitational wave interferometers have provided the first reliable measurement of the rate of coalescing black holes in the local universe, which occur at a rate ranging from 12 to 213 \mbox{yr$^{-1}$ Gpc$^{-3}$} for BHs with masses larger than \mbox{5 $M_\odot$} \citep{Abbott2016c,Abbott2017a}.

After the discovery of such systems, various authors have studied their possible origin \citep[e.g.][]{Dominik2015, Eldridge2016, Marchant2016, Belczynski2016b, Belczynski2017, Kushnir2016, vdHeuvel2017, Hotokezaka2017, Mapelli2017, Bogomazov2018}, concluding that they result for the majority of them from the evolution of isolated massive star binaries born in low metallicity environments. 
In such systems, the stellar evolution quickly leads to the formation of a system of two BHs, that will merge after a long time.
Since the orbital decay time is expected to be much longer than the lifetime of the parent stars, the GW signals point to pairs of black holes which were eventually born several Gyr in the past, and the local rate of binary mergers is the convolution of the history of BBH formation with their coalescence time.
It is also possible to form BBH mergers dynamically in dense stellar clusters, however at smaller rates \citep[e.g.][]{Rodriguez2016}. 

Gamma-ray bursts (GRBs) offer another view at the birth of stellar black holes, especially long GRBs, which are believed to be associated with the core collapse of massive stars  \citep{Hjorth2003, Meszaros2003, Price2003, Vedrenne2009}.
The measure of hundreds of GRB redshifts, thanks to their fast localization with \textit{Swift} \citep{Gehrels2004} and rapid ground based spectroscopic follow-up observations, have led to a good understanding of the history of GRB formation over the ages \citep[e.g.][]{Daigne2006, Wanderman2010, Salvaterra2007, Salvaterra2012, Shahmoradi2013, Howell2014, Lien2014, Petrosian2015, Tan2015, Deng2016, Pescalli2016}.

GRBs and BBH mergers thus offer two complementary views on the history of stellar BH formation, based on two a priori different BH sub-populations. 
This paper aims at comparing the rates of BH formation measured in these two types of events.
Unlike several recent papers, which discuss the emission of gamma-ray bursts during the merger itself \citep[e.g.][]{Connaughton2016, Savchenko2016}, we compare the formation rates of BHs in the past, at the time of their birth. 
Our analysis is based on a simple analytical model which takes into account the current rate of BBH mergers, the time between the formation of binary black holes and their coalescence, the history of GRB formation, the GRB beaming factor, and the possibility of different formation histories of GRBs and BH mergers parametrized with a density evolution index.
This model is described in the next section. 
The comparison of the BH birth rate from BBH mergers and from long GRBs is discussed in Section \ref{sec_results}, while Section \ref{sec_discussion}
adresses some astrophysical consequences of our analysis. 
Cosmological calculations are performed with the astropy.cosmology package with a flat $\Lambda$CDM model with H$_0$~=~\mbox{70 km s$^{-1}$ Mpc$^{-1}$} and $\Omega_{\rm M}$~=~0.3.

%________________________________________________________________

\section{Comparing the rates of BBH mergers and long GRBs}
\label{sec_model}

As discussed in the next section, the coalescence time of a system of binary black holes (the time between the formation of the second BH and the coalescence of the system) strongly depends on the initial separation of the black holes and on their mass. 
According to these parameters, the mergers detected in the local universe are born at different times in the past.
Assuming a power law distribution of coalescence times, the rate of present-day mergers can be written as:

\begin{equation}
\label{eq_merger1}
%\notag
{\mathcal N_{merger}(z=0)} = 0.5 \times {{\int_0^{z_{max}} {\rm N_{\rm BBH}(z)}\ {\rm T_c^{\alpha}(z)}\ d(z)} \over {\int_0^{z_{max}} {\rm T_c^{\alpha}(z)}\ d(z)}} 
\end{equation}

where N$_{\rm BBH}$(z) is the BH birth rate in BBH mergers at redshift z, $\rm T_c$ is the coalescence time of a BBH system born at redshift z and merging at z=0 ($\rm T_c$ is equivalent to the look back time at redshift z), and $\alpha<0$ is the distribution of coalescence times, which depends on the unknown distribution of initial parameters of the BBH system.
The factor 0.5 is justified by the fact that the mergers detected by Advanced LIGO involve two BHs, and $\rm z_{max}$ is the highest redshift at which a BBH contributes to present-day mergers.

Writing $\rm {N_{\rm BBH}(z) = R(z) \times N_{\rm GRB}(z) = R_0 \times \xi(z) \times N_{\rm GRB}(z)}$, where R(z) is the ratio of the number of BHs in BBHs divided by  the number of BHs in GRBs, we can now express equation \ref{eq_merger1} as a function of N$_{\rm GRB}$(z).

\begin{equation}
\label{eq_merger2}
%\notag
{\mathcal N_{merger}} = 0.5 \times {{\int_0^{z_{max}} {\rm R_0}\ \xi(z)\ {\rm N_{\rm GRB}(z)}\ {\rm T_c^{\alpha}(z)}\ d(z)} \over {\int_0^{z_{max}} {\rm T_c^{\alpha}(z)}\ d(z)}}
\end{equation}

$\rm{N_{GRB}(z)}$ can be expressed as a function of the local observed GRB rate $\eta_0$, the beaming factor $\rm f_b$ and the GRB formation history, with the following equation:
\begin{equation}
\label{eq_grb}
%\notag
{\rm N_{GRB}(z) = f_b \times \eta_0  \times \Psi_{GRB}(z)}
\end{equation}

where $\rm \Psi_{GRB}(z)$ follows the history of the comoving GRB formation rate, 
${\rm \Psi_{GRB}(z) =  \big\{ {N_{GRB}(z) \over N_{GRB}(z_0)} \big\}}$.

By replacing $\rm{N_{GRB}(z)}$ in equation \ref{eq_merger2}, we obtain:

\begin{equation}
\label{eq_merger3}
%\notag
{\mathcal N_{merger}} = 0.5 \times R_0 \times f_b \times \eta_0 \times {{\int_0^{z_{max}} \xi(z)\ \Psi_{GRB}(z)\ {\rm T_c^{\alpha}(z)}\ d(z)} \over {\int_0^{z_{max}} {\rm T_c^{\alpha}(z)}\ d(z)}}
\end{equation}

Finally, calling $\rm A(\alpha)$ the ratio of the two intergrals, we compute R as:

\begin{equation}
\label{eq_R}
{\mathcal R(\alpha)} = {\rm {2 \times N_{merger}(z=0) \over f_b \times \eta_0 \times A(\alpha)}}
\end{equation}

This equation underlines the main sources of uncertainty in the evaluation of $\mathcal R $.
One of them is $\rm{N_{merger}(z=0)}$, the local rate of BBH mergers. 
After the detection of four GWs, \cite{Abbott2017a} estimate this rate to be in the range from 12 to 213 \mbox{yr$^{-1}$ Gpc$^{-3}$}, for BHs with masses larger than \mbox{5 $M_\odot$}.
The main reason for this large uncertainty is the small number of detected GW events. 
It is nevertheless expected that this uncertainty will decrease quickly in the future after the detection of more BBH mergers.
In the following, we compute $\mathcal R $ for the two extreme values of $\rm{N_{merger}(z=0)}$.

The beaming factor $\rm f_b = {4\pi \over \Omega}$ (where $\Omega$ is the solid angle of GRB emission) is another important source of uncertainty.
However, since $\mathcal R $ is inversely proportional to $\rm f_b$, it is straightforward to derive $\mathcal R $ for other values of $\rm f_b$.
We adopt $\rm f_b = 250$ in this paper, corresponding to a beaming angle of $\sim$ 5\degr\ \citep{Frail2001, Racusin2009, Ryan2015}.
For comparison, using $\rm f_b = 75$ proposed by \citet{Guetta2005} leads to values of $\mathcal R $ which are more than three times larger.

The distribution of the coalescence times between the formation of the black holes and their coalescence is also uncertain, and we compute $\mathcal R (\alpha)$ for various values of $\alpha$, we have also checked that $\mathcal R $ is independent of $\rm z_{max}$ for $\rm z_{max} \ge 4$.

The other factors have better measurements; this is the case of the local GRB rate $\eta_0$ and of the GRB formation history $\rm \Psi_{GRB}(z)$.
Various authors have studied the GRB formation history \citep[e.g.][]{Daigne2006, Wanderman2010, Salvaterra2007, Salvaterra2012, Shahmoradi2013, Howell2014, Lien2014, Petrosian2015, Tan2015, Deng2016, Pescalli2016}. 
We have chosen here two recent models, which have been validated with \textit{Swift} data and are representative of current estimates of the GRB formation rate. 
The first model, proposed by \citet{Lien2014} \citep[see also][]{Wanderman2010}, uses a broken power law cosmic GRB rate and a broken power law luminosity function whose parameters are adjusted to reproduce the GRB population detected by \textit{Swift}/BAT (see their Table 2). 
In this model, the cosmic GRB formation rate follows the cosmic star formation rate.
The second model, proposed by \cite{Salvaterra2012}, uses a cosmic GRB rate that follows the star formation rate with a density evolution proportional to $\rm (1+z)^{1.7}$, and a broken power law luminosity function (see their Table 2).

The prevalent models of BBH mergers favor their origin from massive stars born in low metallicity regions \citep[e.g.][]{Eldridge2016, Marchant2016, Belczynski2017, Hotokezaka2017} raising the possibility of a different evolution of massive BBHs and GRBs.
We take this possibility into account with a redshift evolution function $\xi(z)$ that favors the creation of massive BBHs at high redshift.
Considering that the formation of GRBs is also favored in low metallicity environments (see Section \ref{sub_grb}), we have chosen evolution parameters that add only moderate extra evolution for BBHs.
We study two functional forms: $\xi(z) = \rm (1+z)^\delta$, and $\xi(z) = 1 / 10^{\beta z}$.
The first option involves a simple parametrization of the density evolution \citep[e.g.][]{Schmidt1999, Salvaterra2012}, for the purpose of this study we have chosen $\delta = 0.6$, which provides a slightly stronger density evolution of BBH with respect to GRBs, by a factor two at redshift z=2.
The second option considers that the formation of BBHs is favored (with respect to the formation of GRBs), in a manner which is inversely proportional to the metallicity. 
For the purpose of this study, we parametrize the evolution of metallicity with $\beta = -0.15$, following the work of \citet{Li2008}.
This function also gives a stronger density evolution of BBHs with respect to GRBs, by a factor two at redshift z=2.

Finally, equation \ref{eq_R} is based on median values, excluding the possibility of correlations between the different parameters.
However, considering the other uncertainties involved in the calculation of $\mathcal R $, this limitation is not decisive in the context of our simplified model.

%and we consider only classical long GRBs (the issue of sub-luminous GRBs is addressed in Section \ref{sec_results}).

%________________________________________________________________
%________________________________________________________________
\begin{figure*}[t]
\centering
\includegraphics[width=0.40\textwidth]{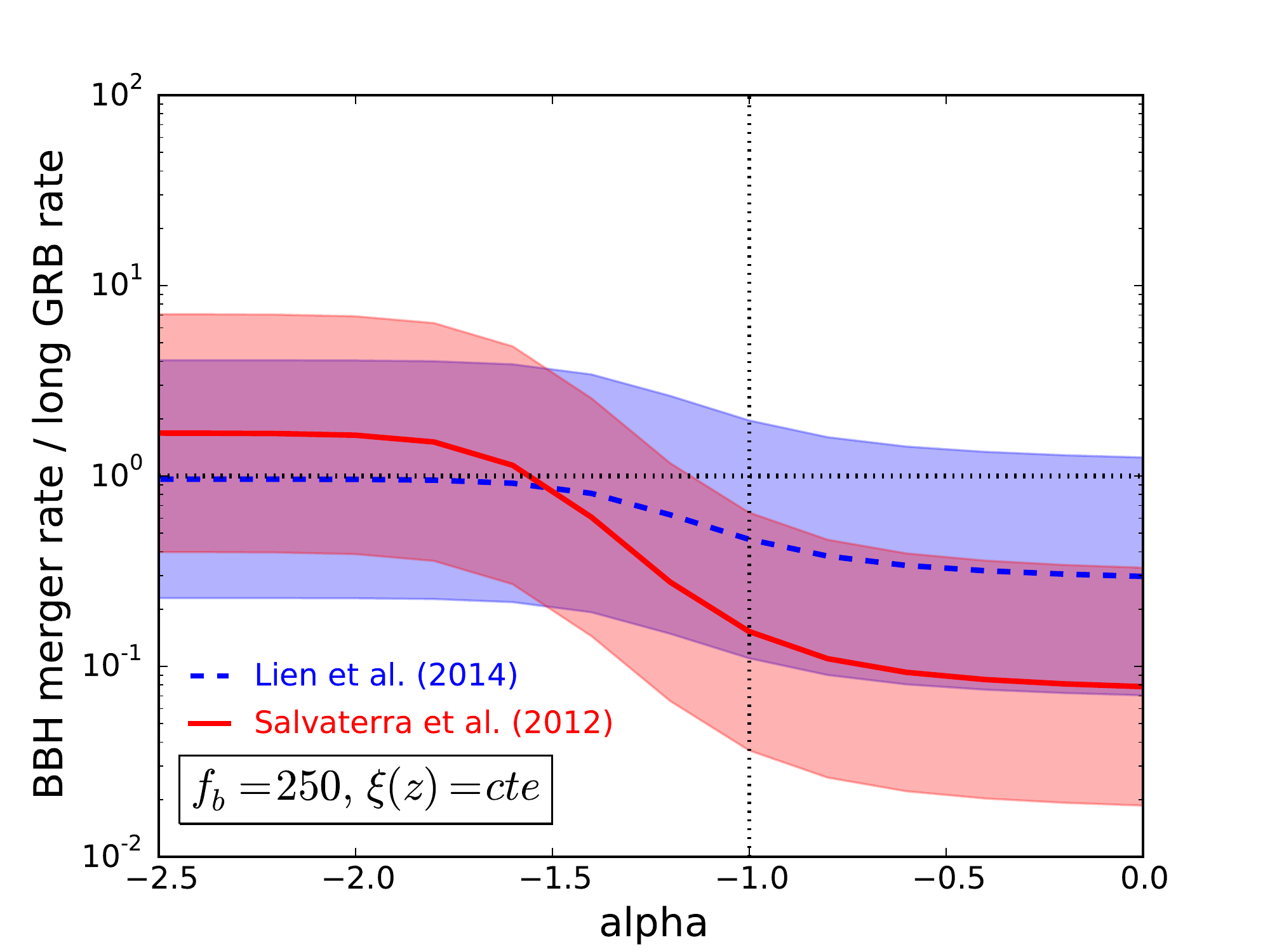}
\includegraphics[width=0.40\textwidth]{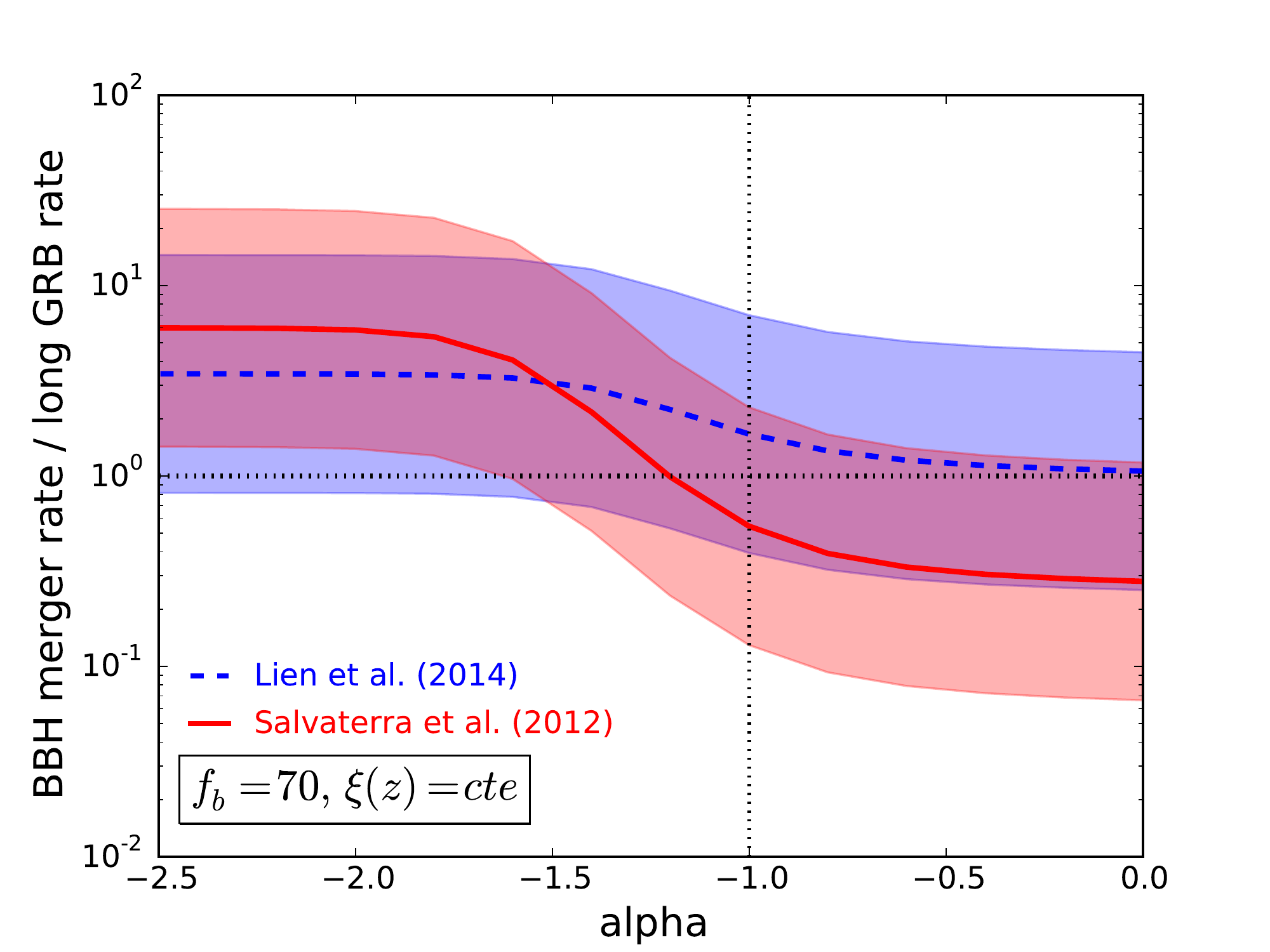}
\includegraphics[width=0.40\textwidth]{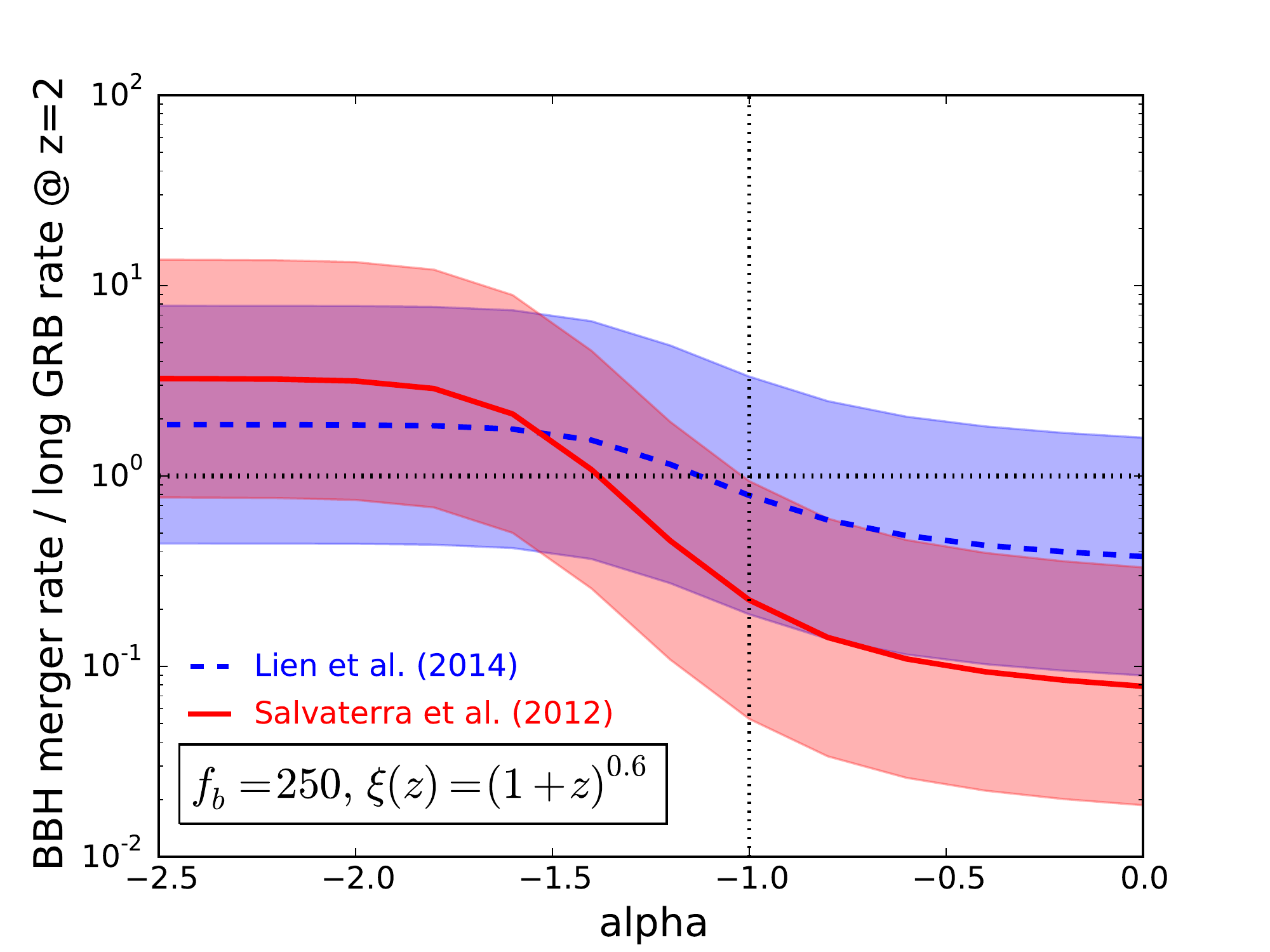}
\includegraphics[width=0.40\textwidth]{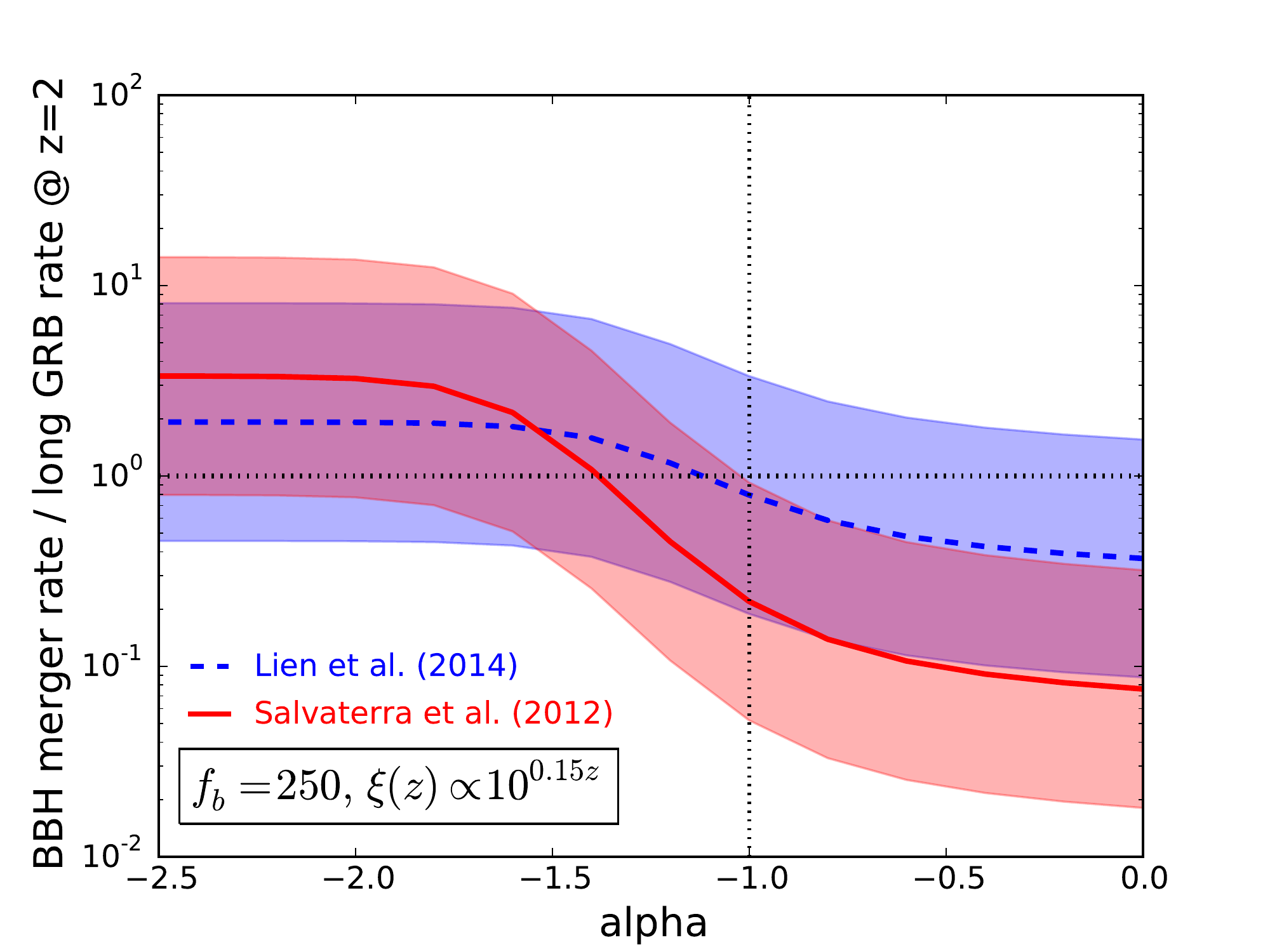}
\caption{The ratio $\mathcal R \rm(\alpha)$ of the stellar BH birth rate measured with BBH mergers and with GRBs, as a function of $\alpha$ the slope of the power law distribution of the coalescence times of present-day mergers. 
The solid and dashed lines show the fraction calculated for a median value of the local rate of BBH mergers measured by Advanced LIGO \citep{Abbott2017a}, while the shaded areas indicate the full range of allowed values considering the errors on the local rate of BBHs mergers.
The parameters indicated in the inset correspond to those appearing in equation \ref{eq_merger3}.
\textbf{Top panels.} $\mathcal R \rm(\alpha)$ is calculated assuming that the rates of GRBs and BBH mergers are proportional, the beaming factor is $\rm f_b = 250$ in the left panel and $\rm f_b = 70$ in the right panel. 
\textbf{Bottom panels.} $\mathcal R \rm(\alpha)$ is calculated assuming that BBH mergers are formed more efficiently than GRBs in the past, the plots show $\mathcal R \rm(\alpha)$ at redshift z=2. The differential density evolution between GRBs and BBH mergers is proportional to $\rm (1+z)^{0.6}$ in the left panel and to $\rm 10^{0.15z}$ in the right panel.
}
\label{fig_ratio}
\end{figure*}

%________________________________________________________________
%________________________________________________________________

\section{Results}
\label{sec_results}
%%%%%% Le rŽsultat principal
Fig. \ref{fig_ratio} shows the value of $\mathcal R (\alpha)$ as a function of $\alpha$.
This figure illustrates the main result of our calculation, which is that, for the fiducial value $\alpha=-1$ \citep[e.g.][]{Belczynski2016b}, the birth rate of stellar black holes in BBH mergers is a few times smaller than the birth rate of stellar black holes in GRBs.
In our reference model, with a beaming factor \mbox{$\rm f_b = 250$} and identical redshift evolution of BH birth in GRBs and BBH mergers (top left panel of  Fig. \ref{fig_ratio}), this ratio varies from few \% to \mbox{$\geq 100$\%}, depending on the assumptions on the history of the GRB formation rate, on $\alpha$, and on the local rate of BBH mergers.
In this model, $\mathcal R $ is in the range [0.1 -- 1.0] for most values of $\alpha$.
The interpretation of this result and some of its astrophysical consequences are briefly addressed in the next section.

From Fig. \ref{fig_ratio} we can derive the main features of $\mathcal R $.
$\mathcal R $ decreases for larger values of $\alpha$, which correspond to an increased contribution of mergers with long coalescence times.
Since BH births were more numerous in the past, the larger contribution of mergers with long coalescence times can produce the present-day merger rate with a smaller fraction of the GRB rate.
$\mathcal R $ is also sensitive to the GRB population model, whose impact is larger for \mbox{$\alpha \ge -1.5$}.
Finally,  $\mathcal R $ is larger for smaller beaming factors, as shown in the top right panel of Fig. \ref{fig_ratio} with \mbox{$\rm f_b = 70$}.
The uncertainty on $\mathcal R $ is expected to decrease quickly in the coming years, with additional detections of BBH mergers, allowing to reduce the size of the shaded areas in Fig.~\ref{fig_ratio}.

\citet{Hotokezaka2017} find that black holes in BBHs and GRBs are produced at roughly similar rates.
We verified that, with the beaming factor \mbox{$\rm f_b = 70$} adopted by these authors, our analysis finds $\mathcal R $ in the range [0.2 -- 10], allowing the formation of BBH mergers and GRBs at about the same rate, like \citet{Hotokezaka2017}. 
This result is illustrated in the top right panel of Fig. \ref{fig_ratio}.

We have not included sub-luminous GRBs in our calculation of $\mathcal R$, since very few have been detected and their link with long classical GRBs is debated.
These events could however change the ratio $\mathcal R $ under two conditions: if sub-luminous GRBs trace the birth of stellar black holes and if their local density is comparable or higher than the local space density of classical GRBs. 
Both conditions are probably met. 
On one hand, the similarity of the supernovae of type Ibc associated with normal and sub-luminous GRBs \citep{Galama1998, Iwamoto1998, Hjorth2003, Stanek2003} justifies the hypothesis that their end products are the same.
On the other hand, various studies have shown that the local rate of sub-luminous GRBs equals or exceeds the rate local of classical GRBs \citep[e.g.][]{Soderberg2004, Daigne2007}.
It is thus quite probable that sub-luminous GRBs represent another population of events associated with the birth of new black holes, this population is however much less known than GRBs or BBHs.

%%%%%% La discussion astro

\section{Discussion}
\label{sec_discussion}

In this section, we discuss the possible connections between BBH mergers and GRBs, in the context of the formation of stellar BHs.

\subsection{Impact of the coalescence time}
\label{sub_tc}
In the calculation of $\mathcal R $, we have considered that present-day mergers have been formed at different epochs in the past, with a distribution of coalescence times following a power-law of slope \mbox{$\alpha < 0$}.
After the birth of the second BH, the decay of the system is governed by general relativity, which predicts that the coalescence time depends on the BH separation to the $4^{\rm th}$ power and to the masses of the BHs to the $3^{\rm rd}$ power \citep{Peters1964}. 
The observation of GW~150914, with a coalescence time smaller than the age of the universe constrains the initial separation between the two BHs to be smaller than $\sim 45 \rm R_\sun$ \citep{Abbott2016b, Hotokezaka2017}, while much smaller separations are required for shorter coalescence times.
Such small separations put strong constraints on the progenitors of BBH mergers, as discussed below.

\subsection{The progenitors of GRBs}
\label{sub_grb}
The connection between GRBs and stellar BH formation encompasses major uncertainties.
While several nearby long GRBs have been associated with exploding massive stars, thanks to their connection with luminous supernovae \citep{Galama1998, Iwamoto1998, Hjorth2003, Stanek2003}, the simulations show that specific conditions are required to produce a GRB (see for instance the reviews of \citet{Woosley2006b} and \citet{Levan2016}). 
First, the core of the exploding star has to keep a large angular momentum, allowing the formation of an accretion disk around the newly formed black hole, a situation that may be facilitated in binary systems \citep[e.g.][]{Podsiadlowski2004, Cantiello2007}.
Second, the relativistic jet must be able to pass through the atmosphere of the progenitor without being slowed down and losing its energy within the star.
This second condition may explain why GRBs have been observed to be associated with supernovae of type Ibc, which have ejected their hydrogen and helium external layers.
Various authors have shown that these conditions, and hence the production of GRBs, are more easily met in low metallicity environments. 
The progenitors of GRBs thus appear to be massive Wolf-Rayet stars born in low metallicity environment, possibly in binary systems. 
Considering these stringent requirements, it is probable that the majority of stellar BHs form silently without a GRB \citep[e.g.][]{MacFadyen1999, Heger2003}. 
This vision is however mitigated by \citet{Levan2016} who note that for a typical beaming factor \mbox{$\rm f_b = 260$}, the long GRB rate is only a factor of three below the formation rate of low metallicity massive stars.

On the other hand, long GRBs may not always be associated with the birth of a BH. 
Indeed, the collapse of a massive star accommodates both the direct collapse to a BH and the formation of a proto-magnetar in those cases where fast-rotating cores produce a magneto-rotational explosion \citep{Dessart2012}. 
The newly born magnetar may be long-lived, or it may eventually collapse to a BH when it spins down. 
From a phenomenological point of view, it has been proposed that the plateau phase observed in ~50\% of long GRBs detected by Swift is an evidence for the presence of a long-lived magnetar central engine \citep{Dai1998, Zhang2001, Corsi2009, Metzger2011, DallOsso2011, Bernardini2012, Bernardini2013}. 
In this context, the sharp drop observed at the end of the plateau in a few cases, which is inconsistent with any afterglow model, has been interpreted as the collapse of an unstable magnetar to a BH \citep{Lyons2010, Rowlinson2013, Chen2017}. 
The fraction of long GRBs that may be powered by a long-lived magnetar could be higher than 50\% \citep[see e.g.][]{Bernardini2015}.

Given these uncertainties, we conclude that the GRB rate cannot be straightforwardly connected with the formation rate of stellar BHs.

\subsection{The progenitors of BBH mergers}
\label{sub_bbh}
Interestingly, the progenitors of BBH mergers appear quite similar to those of GRBs.
The large BH masses and their coalescence time shorter than the age of the universe require massive progenitors born in low metallicity environment, as shown by various authors \citep[e.g.][]{Eldridge2016, Marchant2016, Belczynski2017}.
Moreover, the low aligned spin of the coalescing black holes constrains the radius of their progenitors since large progenitors in close binaries synchronize quickly, leading to a large aligned spin.
This practically reduce the candidates to Wolf-Rayet stars or population III stars \citep[e.g.][]{Kushnir2016, Hotokezaka2017}.

While gravitational wave observations clarify the nature of the progenitors of BBH mergers, at present they are too few to significantly constrain the fraction of stellar BHs that end up in BBH mergers.
While there is an abundant literature on the relative rates of various systems of double compacts objets, we have found few estimates of the fraction of BHs that end their life in BBH mergers. 
Such an estimate is provided by \citet{Elbert2017} who show that the detection rate of BBH mergers with advanced LIGO can be explained if $\sim1\%$ of the total BH population is found in BBH mergers. 
With the detection of an increasing number of BH mergers and the progress of stellar evolution models, it is reasonable to expect better constraints on the fraction of stellar BHs in BBH mergers in the near future.

\subsection{Discussion}
\label{sub_discu}

GRBs and BBH mergers open two windows on the birth of stellar mass black holes.
It is striking to realize that the formation of their progenitors invoke the same ingredients: a massive star born in a low metallicity environment which has lost its external layers. 
Binarity, which is mandatory in BBH mergers, may also play an important role in GRB models. 
At this stage however, we have no observational clue on the possible links between BBH mergers and GRBs, and we briefly speculate on the connection between these two populations with two extreme hypotheses: BHs in BBH mergers are born with a GRB or BHs in BBH mergers are born ``silently'', without a GRB. 

With \mbox{$\mathcal R \leq 1$} for \mbox{$\alpha \ge -1.5$}, our simple analysis shows that the birth rate of BHs in long GRBs is most probably larger than the birth rate of BHs in BBH mergers.
In principle this allows all BHs born in BBH mergers to emit a GRB at birth.
Furthermore, BHs in BBH mergers have more chance than other BHs to emit a GRB at birth since they seem to have similar progenitors. 
Considering that \textit{Swift} has already detected more than 1100 GRBs, several of them could be associated with the birth of BHs in BBH systems.
If we could recognize the GRBs associated with the birth of these BHs, we would be in the situation of observing the beginning and the end of life of binary systems of massive stellar black holes and we could get precious information on the demography and mass spectrum of stellar BHs, on the fraction of BHs born in binary systems, or on the coalescence time.
Unfortunately, for the moment we have no indication of characteristic features that could help distinguishing GRBs associated with the birth of massive BHs in binary systems.

Alternatively, it is possible that BHs born in BBH mergers were born silently (in gamma-rays), making them a distinct population from long GRBs.
Since GRBs require stellar cores with a large angular momentum \citep{MacFadyen1999, Heger2003}, this situation could happen if the progenitors of BBHs are massive stellar cores with low angular momentum, collapsing silently without making a GRB.
If they constitute an entirely distinct population, the BBH mergers may significantly change our vision of the BH demography based on the GRB formation history.

While BH are in principle simple objects uniquely characterized by their mass, charge and their spin, their environment and history play a major role in their observability and in the true distribution of their properties.
It is thus essential to measure the parameters of existing BHs and to understand the biases that affect these measurements if we want to check the validity of theoretical predictions concerning the population of stellar BHs \citep[e.g.][]{Fryer2001, Heger2003, Spera2015}.
In this context, BBH mergers offer new insight into BH formation with accurate mass measurements and information on the BH spins. 
This should help deciphering the formation channel of BBHs and clarify their connection with GRBs and the impact of the BH spin in the formation of GRB jets. 
This could eventually lead to some improvements in our understanding of accretion/ejection processes at work around BHs of all masses and on the formation of relativistic jets.  
In this context, the measurement of the ratio $\mathcal R $ of BH formation in these two subpopulations provides an interesting constraint on future models aiming to simulate the diversity of BH formation channels. 
This ratio can be directly compared with the output of stellar evolution models, as long as they are designed to produce both BBH mergers and GRBs.

In the coming years, the expected increase of BBH and BHNS merger detections \citep{Abbott2016b} and the refinement of merger models will permit a much more precise evaluation of the stellar BH birth rate across the ages, and a much better understanding of the BH population sampled with this technique. 
The increased reach of gravitational waves detectors will also permit measuring the past history of BBH merger formation offering a powerful tool to compare the BH subpopulations explored with BBH mergers and GRBs.
The new understanding coming from GW detections may also lead to a new interpretation of previously known sources. 
\cite{Inoue2016} for instance, propose that Ultra Luminous X-ray sources could be due to a massive star orbiting a young BH, a system that will evolve into a pair of black holes after the explosion of the massive star, and finally into a BBH merger.
A list of several sources which are potential BBH progenitors is also presented in \cite{Bogomazov2018}.
Finally, additional probes of young stellar black hole populations, like sub-luminous GRBs and off-axis GRBs, will soon come into play.
In a few years, the ECLAIRs gamma-ray imager onboard SVOM will explore the real of sub-luminous GRBs with its energy threshold at \mbox{4 \rm keV} \citep{Godet2014, Cordier2015, Wei2016}, while the LSST will detect dozens of optical afterglows from off-axis GRBs in the local universe (see the LSST Science book by the \cite{LSST2009}).

\section{Conclusion}
\label{sec_conclusion}
We have compared the birth rate of stellar black holes measured with two completely different observing channels: BBH mergers detected with GW interferometers and GRBs detected with gamma-ray satellites.
We have taken into account the coalescence time of BBH mergers to perform the comparison at the time of the BH formation. 
The main result of our study is that the birth rate of stellar black holes in BBH mergers is a few times smaller than the rate of long GRBs.
However, considering the numerous uncertainties involved in the calculation, especially on the present rate of BBH mergers and on the beaming factor of GRBs, we cannot exclude that BHs are formed at the same rate in GRBs and in BBH mergers.
This observation raises two questions: what is the fraction of the total BH population in GRBs and in BBH mergers? and what are the possible connections between these two BH subpopulations?
The second question is particularly relevant since GRBs and BBH mergers require similar progenitors: massive stars born in a low metallicity environment.
We suggest that this question could be primarily addressed with stellar evolution models, which should be designed to produce both BBH mergers and GRBs. 
In this context, the measured ratio of the rate of BBH mergers over GRBs provides a constraint that is directly comparable with the predictions of the models.

Finally, the availability of two windows on the population of stellar black holes is only the beginning of the story.
Significant observational progress is expected in the near future, with the increased sensitivity of gravitational wave interferometers, and with the opening of new windows on the birth of stellar BHs in the local universe, like the detection of off-axis GRBs with the LSST and of sub-luminous GRBs with SVOM.

\begin{acknowledgements}
MGB acknowledges the support of the OCEVU Labex (ANR-11-LABX-0060) and the A*MIDEX project (ANR-11-IDEX-0001-02) funded by the ``Investissements d'Avenir'' French government program managed by the ANR.
The authors thank the referee for pointing the constraints imposed by the observation of small aligned spins in binary BH mergers detected with LIGO.
JLA thanks R. Mochkovitch for fruitful discussions that improved the content of the paper.
\end{acknowledgements}


\bibliographystyle{aa} % style aa.bst
\bibliography{biblio-JLA2017} % your references Yourfile.bib

\begin{thebibliography}{71}
\expandafter\ifx\csname natexlab\endcsname\relax\def\natexlab#1{#1}\fi

\bibitem[{{Abbott} {et~al.}(2016{\natexlab{a}}){Abbott}, {Abbott}, {Abbott},
  {Abernathy}, {Acernese}, {Ackley}, {Adams}, {Adams}, {Addesso}, {Adhikari},
  \& et~al.}]{Abbott2016b}
{Abbott}, B.~P., {Abbott}, R., {Abbott}, T.~D., {et~al.} 2016{\natexlab{a}},
  \apjl, 818, L22

\bibitem[{{Abbott} {et~al.}(2016{\natexlab{b}}){Abbott}, {Abbott}, {Abbott},
  {Abernathy}, {Acernese}, {Ackley}, {Adams}, {Adams}, {Addesso}, {Adhikari},
  \& et~al.}]{Abbott2016c}
{Abbott}, B.~P., {Abbott}, R., {Abbott}, T.~D., {et~al.} 2016{\natexlab{b}},
  Physical Review Letters, 116, 241103

\bibitem[{{Abbott} {et~al.}(2016{\natexlab{c}}){Abbott}, {Abbott}, {Abbott},
  {Abernathy}, {Acernese}, {Ackley}, {Adams}, {Adams}, {Addesso}, {Adhikari},
  \& et~al.}]{Abbott2016a}
{Abbott}, B.~P., {Abbott}, R., {Abbott}, T.~D., {et~al.} 2016{\natexlab{c}},
  Physical Review Letters, 116, 061102

\bibitem[{{Abbott} {et~al.}(2017){Abbott}, {Abbott}, {Abbott}, {Acernese},
  {Ackley}, {Adams}, {Adams}, {Addesso}, {Adhikari}, {Adya}, \&
  et~al.}]{Abbott2017a}
{Abbott}, B.~P., {Abbott}, R., {Abbott}, T.~D., {et~al.} 2017, Physical Review
  Letters, 118, 221101

\bibitem[{{Belczynski} {et~al.}(2016){Belczynski}, {Holz}, {Bulik}, \&
  {O'Shaughnessy}}]{Belczynski2016b}
{Belczynski}, K., {Holz}, D.~E., {Bulik}, T., \& {O'Shaughnessy}, R. 2016,
  \nat, 534, 512

\bibitem[{{Belczynski} {et~al.}(2017){Belczynski}, {Klencki}, {Meynet},
  {Fryer}, {Brown}, {Chruslinska}, {Gladysz}, {O'Shaughnessy}, {Bulik},
  {Berti}, {Holz}, {Gerosa}, {Giersz}, {Ekstrom}, {Georgy}, {Askar}, \&
  {Lasota}}]{Belczynski2017}
{Belczynski}, K., {Klencki}, J., {Meynet}, G., {et~al.} 2017, ArXiv e-prints
  [\eprint[arXiv]{1706.07053}]

\bibitem[{{Bernardini}(2015)}]{Bernardini2015}
{Bernardini}, M.~G. 2015, Journal of High Energy Astrophysics, 7, 64

\bibitem[{{Bernardini} {et~al.}(2013){Bernardini}, {Campana}, {Ghisellini},
  {D'Avanzo}, {Burlon}, {Covino}, {Ghirlanda}, {Melandri}, {Salvaterra},
  {Vergani}, {D'Elia}, {Fugazza}, {Sbarufatti}, \&
  {Tagliaferri}}]{Bernardini2013}
{Bernardini}, M.~G., {Campana}, S., {Ghisellini}, G., {et~al.} 2013, \apj, 775,
  67

\bibitem[{{Bernardini} {et~al.}(2012){Bernardini}, {Margutti}, {Mao},
  {Zaninoni}, \& {Chincarini}}]{Bernardini2012}
{Bernardini}, M.~G., {Margutti}, R., {Mao}, J., {Zaninoni}, E., \&
  {Chincarini}, G. 2012, \aap, 539, A3

\bibitem[{{Bogomazov} {et~al.}(2018){Bogomazov}, {Cherepashchuk}, {Lipunov}, \&
  {Tutukov}}]{Bogomazov2018}
{Bogomazov}, A.~I., {Cherepashchuk}, A.~M., {Lipunov}, V.~M., \& {Tutukov},
  A.~V. 2018, \na, 58, 33

\bibitem[{{Cantiello} {et~al.}(2007){Cantiello}, {Yoon}, {Langer}, \&
  {Livio}}]{Cantiello2007}
{Cantiello}, M., {Yoon}, S.-C., {Langer}, N., \& {Livio}, M. 2007, \aap, 465,
  L29

\bibitem[{{Chen} {et~al.}(2017){Chen}, {Xie}, {Lei}, {Zou}, {L{\"u}}, {Liang},
  {Gao}, \& {Wang}}]{Chen2017}
{Chen}, W., {Xie}, W., {Lei}, W.-H., {et~al.} 2017, ArXiv e-prints
  [\eprint[arXiv]{1709.08285}]

\bibitem[{{Connaughton} {et~al.}(2016){Connaughton}, {Burns}, {Goldstein},
  {Blackburn}, {Briggs}, {Zhang}, {Camp}, {Christensen}, {Hui}, {Jenke},
  {Littenberg}, {McEnery}, {Racusin}, {Shawhan}, {Singer}, {Veitch},
  {Wilson-Hodge}, {Bhat}, {Bissaldi}, {Cleveland}, {Fitzpatrick}, {Giles},
  {Gibby}, {von Kienlin}, {Kippen}, {McBreen}, {Mailyan}, {Meegan}, {Paciesas},
  {Preece}, {Roberts}, {Sparke}, {Stanbro}, {Toelge}, \&
  {Veres}}]{Connaughton2016}
{Connaughton}, V., {Burns}, E., {Goldstein}, A., {et~al.} 2016, \apjl, 826, L6

\bibitem[{{Cordier} {et~al.}(2015){Cordier}, {Wei}, {Atteia}, {Basa}, {Claret},
  {Daigne}, {Deng}, {Dong}, {Godet}, {Goldwurm}, {G{\"o}tz}, {Han}, {Klotz},
  {Lachaud}, {Osborne}, {Qiu}, {Schanne}, {Wu}, {Wang}, {Wu}, {Xin}, {Zhang},
  \& {Zhang}}]{Cordier2015}
{Cordier}, B., {Wei}, J., {Atteia}, J.-L., {et~al.} 2015, ArXiv e-prints
  [\eprint[arXiv]{1512.03323}]

\bibitem[{{Corsi} \& {M{\'e}sz{\'a}ros}(2009)}]{Corsi2009}
{Corsi}, A. \& {M{\'e}sz{\'a}ros}, P. 2009, \apj, 702, 1171

\bibitem[{{Dai} \& {Lu}(1998)}]{Dai1998}
{Dai}, Z.~G. \& {Lu}, T. 1998, Physical Review Letters, 81, 4301

\bibitem[{{Daigne} \& {Mochkovitch}(2007)}]{Daigne2007}
{Daigne}, F. \& {Mochkovitch}, R. 2007, \aap, 465, 1

\bibitem[{{Daigne} {et~al.}(2006){Daigne}, {Rossi}, \&
  {Mochkovitch}}]{Daigne2006}
{Daigne}, F., {Rossi}, E.~M., \& {Mochkovitch}, R. 2006, \mnras, 372, 1034

\bibitem[{{Dall'Osso} {et~al.}(2011){Dall'Osso}, {Stratta}, {Guetta}, {Covino},
  {De Cesare}, \& {Stella}}]{DallOsso2011}
{Dall'Osso}, S., {Stratta}, G., {Guetta}, D., {et~al.} 2011, \aap, 526, A121

\bibitem[{{Deng} {et~al.}(2016){Deng}, {Wang}, {Guo}, {Lu}, {Wang}, {Wei},
  {Wu}, \& {Liang}}]{Deng2016}
{Deng}, C.-M., {Wang}, X.-G., {Guo}, B.-B., {et~al.} 2016, \apj, 820, 66

\bibitem[{{Dessart} {et~al.}(2012){Dessart}, {O'Connor}, \&
  {Ott}}]{Dessart2012}
{Dessart}, L., {O'Connor}, E., \& {Ott}, C.~D. 2012, \apj, 754, 76

\bibitem[{{Dominik} {et~al.}(2015){Dominik}, {Berti}, {O'Shaughnessy},
  {Mandel}, {Belczynski}, {Fryer}, {Holz}, {Bulik}, \&
  {Pannarale}}]{Dominik2015}
{Dominik}, M., {Berti}, E., {O'Shaughnessy}, R., {et~al.} 2015, \apj, 806, 263

\bibitem[{{Elbert} {et~al.}(2017){Elbert}, {Bullock}, \&
  {Kaplinghat}}]{Elbert2017}
{Elbert}, O.~D., {Bullock}, J.~S., \& {Kaplinghat}, M. 2017, ArXiv e-prints
  [\eprint[arXiv]{1703.02551}]

\bibitem[{{Eldridge} \& {Stanway}(2016)}]{Eldridge2016}
{Eldridge}, J.~J. \& {Stanway}, E.~R. 2016, \mnras, 462, 3302

\bibitem[{{Frail} {et~al.}(2001){Frail}, {Kulkarni}, {Sari}, {Djorgovski},
  {Bloom}, {Galama}, {Reichart}, {Berger}, {Harrison}, {Price}, {Yost},
  {Diercks}, {Goodrich}, \& {Chaffee}}]{Frail2001}
{Frail}, D.~A., {Kulkarni}, S.~R., {Sari}, R., {et~al.} 2001, \apjl, 562, L55

\bibitem[{{Fryer} \& {Kalogera}(2001)}]{Fryer2001}
{Fryer}, C.~L. \& {Kalogera}, V. 2001, \apj, 554, 548

\bibitem[{{Galama} {et~al.}(1998){Galama}, {Vreeswijk}, {van Paradijs},
  {Kouveliotou}, {Augusteijn}, {B{\"o}hnhardt}, {Brewer}, {Doublier},
  {Gonzalez}, {Leibundgut}, {Lidman}, {Hainaut}, {Patat}, {Heise}, {in't Zand},
  {Hurley}, {Groot}, {Strom}, {Mazzali}, {Iwamoto}, {Nomoto}, {Umeda},
  {Nakamura}, {Young}, {Suzuki}, {Shigeyama}, {Koshut}, {Kippen}, {Robinson},
  {de Wildt}, {Wijers}, {Tanvir}, {Greiner}, {Pian}, {Palazzi}, {Frontera},
  {Masetti}, {Nicastro}, {Feroci}, {Costa}, {Piro}, {Peterson}, {Tinney},
  {Boyle}, {Cannon}, {Stathakis}, {Sadler}, {Begam}, \& {Ianna}}]{Galama1998}
{Galama}, T.~J., {Vreeswijk}, P.~M., {van Paradijs}, J., {et~al.} 1998, \nat,
  395, 670

\bibitem[{{Gehrels} {et~al.}(2004){Gehrels}, {Chincarini}, {Giommi}, {Mason},
  {Nousek}, {Wells}, {White}, {Barthelmy}, {Burrows}, {Cominsky}, {Hurley},
  {Marshall}, {M{\'e}sz{\'a}ros}, {Roming}, {Angelini}, {Barbier}, {Belloni},
  {Campana}, {Caraveo}, {Chester}, {Citterio}, {Cline}, {Cropper}, {Cummings},
  {Dean}, {Feigelson}, {Fenimore}, {Frail}, {Fruchter}, {Garmire}, {Gendreau},
  {Ghisellini}, {Greiner}, {Hill}, {Hunsberger}, {Krimm}, {Kulkarni}, {Kumar},
  {Lebrun}, {Lloyd-Ronning}, {Markwardt}, {Mattson}, {Mushotzky}, {Norris},
  {Osborne}, {Paczynski}, {Palmer}, {Park}, {Parsons}, {Paul}, {Rees},
  {Reynolds}, {Rhoads}, {Sasseen}, {Schaefer}, {Short}, {Smale}, {Smith},
  {Stella}, {Tagliaferri}, {Takahashi}, {Tashiro}, {Townsley}, {Tueller},
  {Turner}, {Vietri}, {Voges}, {Ward}, {Willingale}, {Zerbi}, \&
  {Zhang}}]{Gehrels2004}
{Gehrels}, N., {Chincarini}, G., {Giommi}, P., {et~al.} 2004, \apj, 611, 1005

\bibitem[{{Godet} {et~al.}(2014){Godet}, {Nasser}, {Atteia}, {Cordier},
  {Mandrou}, {Barret}, {Triou}, {Pons}, {Amoros}, {Bordon}, {Gevin},
  {Gonzalez}, {G{\"o}tz}, {Gros}, {Houret}, {Lachaud}, {Lacombe}, {Marty},
  {Mercier}, {Rambaud}, {Ramon}, {Rouaix}, {Schanne}, \&
  {Waegebaert}}]{Godet2014}
{Godet}, O., {Nasser}, G., {Atteia}, J.-., {et~al.} 2014, in \procspie, Vol.
  9144, Space Telescopes and Instrumentation 2014: Ultraviolet to Gamma Ray,
  914424

\bibitem[{{Guetta} {et~al.}(2005){Guetta}, {Piran}, \& {Waxman}}]{Guetta2005}
{Guetta}, D., {Piran}, T., \& {Waxman}, E. 2005, \apj, 619, 412

\bibitem[{{Heger} {et~al.}(2003){Heger}, {Fryer}, {Woosley}, {Langer}, \&
  {Hartmann}}]{Heger2003}
{Heger}, A., {Fryer}, C.~L., {Woosley}, S.~E., {Langer}, N., \& {Hartmann},
  D.~H. 2003, \apj, 591, 288

\bibitem[{{Hjorth} {et~al.}(2003){Hjorth}, {Sollerman}, {M{\o}ller}, {Fynbo},
  {Woosley}, {Kouveliotou}, {Tanvir}, {Greiner}, {Andersen}, {Castro-Tirado},
  {Castro Cer{\'o}n}, {Fruchter}, {Gorosabel}, {Jakobsson}, {Kaper}, {Klose},
  {Masetti}, {Pedersen}, {Pedersen}, {Pian}, {Palazzi}, {Rhoads}, {Rol}, {van
  den Heuvel}, {Vreeswijk}, {Watson}, \& {Wijers}}]{Hjorth2003}
{Hjorth}, J., {Sollerman}, J., {M{\o}ller}, P., {et~al.} 2003, \nat, 423, 847

\bibitem[{{Hotokezaka} \& {Piran}(2017)}]{Hotokezaka2017}
{Hotokezaka}, K. \& {Piran}, T. 2017, \apj, 842, 111

\bibitem[{{Howell} {et~al.}(2014){Howell}, {Coward}, {Stratta}, {Gendre}, \&
  {Zhou}}]{Howell2014}
{Howell}, E.~J., {Coward}, D.~M., {Stratta}, G., {Gendre}, B., \& {Zhou}, H.
  2014, \mnras, 444, 15

\bibitem[{{Inoue} {et~al.}(2016){Inoue}, {Tanaka}, \& {Isobe}}]{Inoue2016}
{Inoue}, Y., {Tanaka}, Y.~T., \& {Isobe}, N. 2016, \mnras, 461, 4329

\bibitem[{{Iwamoto} {et~al.}(1998){Iwamoto}, {Mazzali}, {Nomoto}, {Umeda},
  {Nakamura}, {Patat}, {Danziger}, {Young}, {Suzuki}, {Shigeyama},
  {Augusteijn}, {Doublier}, {Gonzalez}, {Boehnhardt}, {Brewer}, {Hainaut},
  {Lidman}, {Leibundgut}, {Cappellaro}, {Turatto}, {Galama}, {Vreeswijk},
  {Kouveliotou}, {van Paradijs}, {Pian}, {Palazzi}, \&
  {Frontera}}]{Iwamoto1998}
{Iwamoto}, K., {Mazzali}, P.~A., {Nomoto}, K., {et~al.} 1998, \nat, 395, 672

\bibitem[{{Kushnir} {et~al.}(2016){Kushnir}, {Zaldarriaga}, {Kollmeier}, \&
  {Waldman}}]{Kushnir2016}
{Kushnir}, D., {Zaldarriaga}, M., {Kollmeier}, J.~A., \& {Waldman}, R. 2016,
  \mnras, 462, 844

\bibitem[{{Levan} {et~al.}(2016){Levan}, {Crowther}, {de Grijs}, {Langer},
  {Xu}, \& {Yoon}}]{Levan2016}
{Levan}, A., {Crowther}, P., {de Grijs}, R., {et~al.} 2016, \ssr, 202, 33

\bibitem[{{Li}(2008)}]{Li2008}
{Li}, L.-X. 2008, \mnras, 388, 1487

\bibitem[{{Lien} {et~al.}(2014){Lien}, {Sakamoto}, {Gehrels}, {Palmer},
  {Barthelmy}, {Graziani}, \& {Cannizzo}}]{Lien2014}
{Lien}, A., {Sakamoto}, T., {Gehrels}, N., {et~al.} 2014, \apj, 783, 24

\bibitem[{{LSST Science Collaboration} {et~al.}(2009){LSST Science
  Collaboration}, {Abell}, {Allison}, {Anderson}, {Andrew}, {Angel}, {Armus},
  {Arnett}, {Asztalos}, {Axelrod}, \& et~al.}]{LSST2009}
{LSST Science Collaboration}, {Abell}, P.~A., {Allison}, J., {et~al.} 2009,
  ArXiv e-prints [\eprint[arXiv]{0912.0201}]

\bibitem[{{Lyons} {et~al.}(2010){Lyons}, {O'Brien}, {Zhang}, {Willingale},
  {Troja}, \& {Starling}}]{Lyons2010}
{Lyons}, N., {O'Brien}, P.~T., {Zhang}, B., {et~al.} 2010, \mnras, 402, 705

\bibitem[{{MacFadyen} \& {Woosley}(1999)}]{MacFadyen1999}
{MacFadyen}, A.~I. \& {Woosley}, S.~E. 1999, \apj, 524, 262

\bibitem[{{Mapelli} {et~al.}(2017){Mapelli}, {Giacobbo}, {Ripamonti}, \&
  {Spera}}]{Mapelli2017}
{Mapelli}, M., {Giacobbo}, N., {Ripamonti}, E., \& {Spera}, M. 2017, ArXiv
  e-prints [\eprint[arXiv]{1708.05722}]

\bibitem[{{Marchant} {et~al.}(2016){Marchant}, {Langer}, {Podsiadlowski},
  {Tauris}, \& {Moriya}}]{Marchant2016}
{Marchant}, P., {Langer}, N., {Podsiadlowski}, P., {Tauris}, T.~M., \&
  {Moriya}, T.~J. 2016, \aap, 588, A50

\bibitem[{{M{\'e}sz{\'a}ros}(2003)}]{Meszaros2003}
{M{\'e}sz{\'a}ros}, P. 2003, \nat, 423, 809

\bibitem[{{Metzger} {et~al.}(2011){Metzger}, {Giannios}, {Thompson},
  {Bucciantini}, \& {Quataert}}]{Metzger2011}
{Metzger}, B.~D., {Giannios}, D., {Thompson}, T.~A., {Bucciantini}, N., \&
  {Quataert}, E. 2011, \mnras, 413, 2031

\bibitem[{{Pescalli} {et~al.}(2016){Pescalli}, {Ghirlanda}, {Salvaterra},
  {Ghisellini}, {Vergani}, {Nappo}, {Salafia}, {Melandri}, {Covino}, \&
  {G{\"o}tz}}]{Pescalli2016}
{Pescalli}, A., {Ghirlanda}, G., {Salvaterra}, R., {et~al.} 2016, \aap, 587,
  A40

\bibitem[{{Peters}(1964)}]{Peters1964}
{Peters}, P.~C. 1964, Physical Review, 136, 1224

\bibitem[{{Petrosian} {et~al.}(2015){Petrosian}, {Kitanidis}, \&
  {Kocevski}}]{Petrosian2015}
{Petrosian}, V., {Kitanidis}, E., \& {Kocevski}, D. 2015, \apj, 806, 44

\bibitem[{{Podsiadlowski} {et~al.}(2004){Podsiadlowski}, {Mazzali}, {Nomoto},
  {Lazzati}, \& {Cappellaro}}]{Podsiadlowski2004}
{Podsiadlowski}, P., {Mazzali}, P.~A., {Nomoto}, K., {Lazzati}, D., \&
  {Cappellaro}, E. 2004, \apjl, 607, L17

\bibitem[{{Price} {et~al.}(2003){Price}, {Fox}, {Kulkarni}, {Peterson},
  {Schmidt}, {Soderberg}, {Yost}, {Berger}, {Djorgovski}, {Frail}, {Harrison},
  {Sari}, {Blain}, \& {Chapman}}]{Price2003}
{Price}, P.~A., {Fox}, D.~W., {Kulkarni}, S.~R., {et~al.} 2003, \nat, 423, 844

\bibitem[{{Racusin} {et~al.}(2009){Racusin}, {Liang}, {Burrows}, {Falcone},
  {Sakamoto}, {Zhang}, {Zhang}, {Evans}, \& {Osborne}}]{Racusin2009}
{Racusin}, J.~L., {Liang}, E.~W., {Burrows}, D.~N., {et~al.} 2009, \apj, 698,
  43

\bibitem[{{Rodriguez} {et~al.}(2016){Rodriguez}, {Chatterjee}, \&
  {Rasio}}]{Rodriguez2016}
{Rodriguez}, C.~L., {Chatterjee}, S., \& {Rasio}, F.~A. 2016, \prd, 93, 084029

\bibitem[{{Rowlinson} {et~al.}(2013){Rowlinson}, {O'Brien}, {Metzger},
  {Tanvir}, \& {Levan}}]{Rowlinson2013}
{Rowlinson}, A., {O'Brien}, P.~T., {Metzger}, B.~D., {Tanvir}, N.~R., \&
  {Levan}, A.~J. 2013, \mnras, 430, 1061

\bibitem[{{Ryan} {et~al.}(2015){Ryan}, {van Eerten}, {MacFadyen}, \&
  {Zhang}}]{Ryan2015}
{Ryan}, G., {van Eerten}, H., {MacFadyen}, A., \& {Zhang}, B.-B. 2015, \apj,
  799, 3

\bibitem[{{Salvaterra} {et~al.}(2012){Salvaterra}, {Campana}, {Vergani},
  {Covino}, {D'Avanzo}, {Fugazza}, {Ghirlanda}, {Ghisellini}, {Melandri},
  {Nava}, {Sbarufatti}, {Flores}, {Piranomonte}, \&
  {Tagliaferri}}]{Salvaterra2012}
{Salvaterra}, R., {Campana}, S., {Vergani}, S.~D., {et~al.} 2012, \apj, 749, 68

\bibitem[{{Salvaterra} \& {Chincarini}(2007)}]{Salvaterra2007}
{Salvaterra}, R. \& {Chincarini}, G. 2007, \apjl, 656, L49

\bibitem[{{Savchenko} {et~al.}(2016){Savchenko}, {Ferrigno}, {Mereghetti},
  {Natalucci}, {Bazzano}, {Bozzo}, {Brandt}, {Courvoisier}, {Diehl}, {Hanlon},
  {von Kienlin}, {Kuulkers}, {Laurent}, {Lebrun}, {Roques}, {Ubertini}, \&
  {Weidenspointner}}]{Savchenko2016}
{Savchenko}, V., {Ferrigno}, C., {Mereghetti}, S., {et~al.} 2016, \apjl, 820,
  L36

\bibitem[{{Schmidt}(1999)}]{Schmidt1999}
{Schmidt}, M. 1999, \apjl, 523, L117

\bibitem[{{Shahmoradi}(2013)}]{Shahmoradi2013}
{Shahmoradi}, A. 2013, \apj, 766, 111

\bibitem[{{Soderberg} {et~al.}(2004){Soderberg}, {Kulkarni}, {Berger}, {Fox},
  {Sako}, {Frail}, {Gal-Yam}, {Moon}, {Cenko}, {Yost}, {Phillips}, {Persson},
  {Freedman}, {Wyatt}, {Jayawardhana}, \& {Paulson}}]{Soderberg2004}
{Soderberg}, A.~M., {Kulkarni}, S.~R., {Berger}, E., {et~al.} 2004, \nat, 430,
  648

\bibitem[{{Spera} {et~al.}(2015){Spera}, {Mapelli}, \& {Bressan}}]{Spera2015}
{Spera}, M., {Mapelli}, M., \& {Bressan}, A. 2015, \mnras, 451, 4086

\bibitem[{{Stanek} {et~al.}(2003){Stanek}, {Matheson}, {Garnavich}, {Martini},
  {Berlind}, {Caldwell}, {Challis}, {Brown}, {Schild}, {Krisciunas}, {Calkins},
  {Lee}, {Hathi}, {Jansen}, {Windhorst}, {Echevarria}, {Eisenstein}, {Pindor},
  {Olszewski}, {Harding}, {Holland}, \& {Bersier}}]{Stanek2003}
{Stanek}, K.~Z., {Matheson}, T., {Garnavich}, P.~M., {et~al.} 2003, \apjl, 591,
  L17

\bibitem[{{Tan} \& {Wang}(2015)}]{Tan2015}
{Tan}, W.-W. \& {Wang}, F.~Y. 2015, \mnras, 454, 1785

\bibitem[{{van den Heuvel} {et~al.}(2017){van den Heuvel}, {Portegies Zwart},
  \& {de Mink}}]{vdHeuvel2017}
{van den Heuvel}, E.~P.~J., {Portegies Zwart}, S.~F., \& {de Mink}, S.~E. 2017,
  ArXiv e-prints [\eprint[arXiv]{1701.02355}]

\bibitem[{Vedrenne \& Atteia(2009)}]{Vedrenne2009}
Vedrenne, G. \& Atteia, J. 2009, Gamma-Ray Bursts: The brightest explosions in
  the Universe, Springer Praxis Books (Springer Berlin Heidelberg)

\bibitem[{{Wanderman} \& {Piran}(2010)}]{Wanderman2010}
{Wanderman}, D. \& {Piran}, T. 2010, \mnras, 406, 1944

\bibitem[{{Wei} {et~al.}(2016){Wei}, {Cordier}, {Antier}, {Antilogus},
  {Atteia}, {Bajat}, {Basa}, {Beckmann}, {Bernardini}, {Boissier}, {Bouchet},
  {Burwitz}, {Claret}, {Dai}, {Daigne}, {Deng}, {Dornic}, {Feng}, {Foglizzo},
  {Gao}, {Gehrels}, {Godet}, {Goldwurm}, {Gonzalez}, {Gosset}, {G{\"o}tz},
  {Gouiffes}, {Grise}, {Gros}, {Guilet}, {Han}, {Huang}, {Huang}, {Jouret},
  {Klotz}, {La Marle}, {Lachaud}, {Le Floch}, {Lee}, {Leroy}, {Li}, {Li}, {Li},
  {Liang}, {Lyu}, {Mercier}, {Migliori}, {Mochkovitch}, {O'Brien}, {Osborne},
  {Paul}, {Perinati}, {Petitjean}, {Piron}, {Qiu}, {Rau}, {Rodriguez},
  {Schanne}, {Tanvir}, {Vangioni}, {Vergani}, {Wang}, {Wang}, {Wang}, {Wang},
  {Watson}, {Webb}, {Wei}, {Willingale}, {Wu}, {Wu}, {Xin}, {Xu}, {Yu}, {Yu},
  {Yu}, {Zhang}, {Zhang}, {Zhang}, \& {Zhou}}]{Wei2016}
{Wei}, J., {Cordier}, B., {Antier}, S., {et~al.} 2016, ArXiv e-prints
  [\eprint[arXiv]{1610.06892}]

\bibitem[{{Woosley} \& {Bloom}(2006)}]{Woosley2006b}
{Woosley}, S.~E. \& {Bloom}, J.~S. 2006, \araa, 44, 507

\bibitem[{{Zhang} \& {M{\'e}sz{\'a}ros}(2001)}]{Zhang2001}
{Zhang}, B. \& {M{\'e}sz{\'a}ros}, P. 2001, \apjl, 552, L35

\end{thebibliography}

\end{document}